\documentclass{aipproc}

\renewcommand{\S}{{\tt S}}  
\renewcommand{\P}{{\tt P}}  

\newcommand{\G}{{\tt G}}

\newcommand{\mGeV}{$\mbox{GeV/c}^2$}		
\newcommand{\pGeV}{$\mbox{GeV/c}$}		
\newcommand{\tGeV}{$\mbox{GeV}^2\!/c^2$}	
\newcommand{\eps}{\varepsilon}

\layoutstyle{6x9}

\begin{document}

\title{Study of reaction $\pi^- A \to \pi^+\pi^-\pi^- A$ at VES setup.}


\author{Igor Kachaev for VES Collaboration.
  \thanks{Amelin~D.V., Dorofeev~V.A., Dzhelyadin~R.I., Gouz~Yu.P.,
  Kachaev~I.A., Karyukhin~A.N., Khokh\-lov~Yu.A.,
  Ko\-no\-plyan\-nikov~A.K., Konstantinov~V.F.,
  Kopikov~S.V., Kostyukhin~V.V., Kostyukhina~I.V.,  Matveev~V.D.,
  Nikolaenko~V.I., Ostankov~A.P., Polyakov~B.F., Ryabchikov~D.I.,
  Solodkov~A.A., Solovianov~O.V., Zaitsev~A.M.}
}
{address={Department of Hadron Physics, IHEP, Protvino, Russia, 142284},
 email={kachaev@mx.ihep.su}}
\copyrightyear{2001}
\date{\today}

\begin{abstract}

  The results on partial wave analysis of $3\pi$ system in reaction
$\pi^- A \to \pi^+\pi^-\pi^- A$
at the momentum $36.6$~\pGeV{} on the beryllium target are presented.
New method of amplitude analysis is suggested --- extraction of largest
eigenvalue of density matrix.
Exotic wave with $J^{PC} = 1^{-+} \rho\pi$ is studied in four $t'$
regions. No narrow object around $M=1.6$~\mGeV{} is found.
Unusually steep $t'$ dependence for $\pi(1300)$ object is detected.

\end{abstract}
\maketitle

\section{Introduction. The Partial Wave Analysis.}

   The VErtex Spectrometer (VES) setup is a large aperture
magnetic spectrometer including the system of proportional and drift chambers, 
a multichannel threshold \v{C}erenkov counter,
beam-line \v{C}erenkov counters,
a lead-glass $\gamma$-detector (LGD) and trigger hodoscope.
This permits full identification of multi-particle final states.
The setup runs on the negative particle beam with the momentum of 36.6~\pGeV.
The description of the setup can be found in~\cite{bec1}.

   In this report we present some results of partial wave analysis
of the $3\pi$ system in the reaction $\pi^- Be \to \pi^+\pi^-\pi^- Be$
for different $t'$ regions
$|t'|=0.01-0.07-0.15-0.30-0.80$ \tGeV.
  The discussed results are based upon the statistics of about
$8.0\cdot10^6$ events.
Our previous results were published in \cite{pippm}.

  The PWA has been performed in the
0.8--2.6~\mGeV{} mass region in 50~MeV bins for different $t'$ regions.
Modified version of the
Illinois PWA program \cite{hansen} with maximum likehood method 
has been used for the analysis.
Amplitudes were written using isobar model and relativistic covariant
helicity formalism according to \cite{ChungAmp}.
Explicit $t'$-dependence $f(t)=te^{-bt}$ was included for waves
with nonzero projection of spin on GJ z-axis.
Density matrix of full rank was used to describe final state.
The set of 42 partial waves in the form
${\rm J^{P}LM^{\textstyle\eta}}\,isobar$ \cite{hansen}
was used in the analysis.
Full wave set and parameterization of isobars including special treatment of
$\pi\pi$ S-wave can be found in \cite{pippm}.
For the channels with $J^{PC}=0^{-+}$, $1^{++}$, $2^{-+}$ largest waves
are assigned to their own density matrix elements
and are enabled to freely interfere with each other. 

\section{Extraction of largest eigenvalue.}

\begin{figure}[tb]
\includegraphics[width=\textwidth]{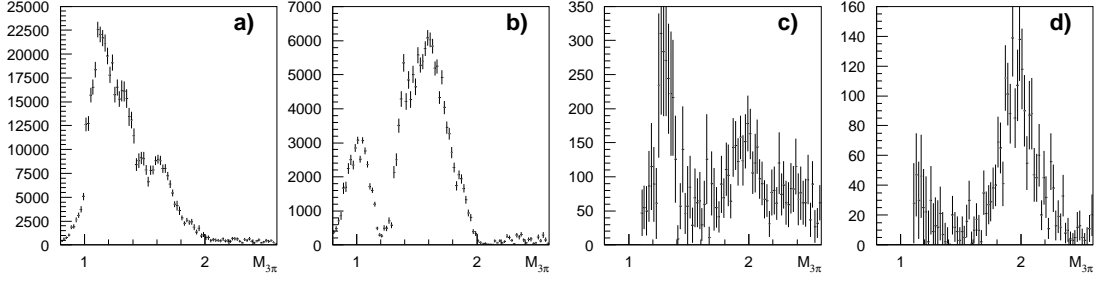}
\caption{Wave $2^-\P0^+\,\rho\pi$ at $|t'|<0.03$~\tGeV{} in
a) full density matrix, b) largest eigenvalue;\protect\\
wave $4^+\G1^+\,\rho\pi$ at $|t'|<0.03$~\tGeV{} in
c) full density matrix, d) largest eigenvalue.}
\label{fig-lev}
\end{figure}

Results of PWA are represented in general by {\it density matrix}.
For physical analysis {\it amplitudes} are much more convenient.
We present here a new type of amplitude analysis --- extraction of largest
eigenvalue of density matrix.
Density matrix can be represented by its eigenvalues and eigenvectors:
\[ \rho = \sum_{k=1}^d e_k*V_k*V_k^+ \quad \mbox{where} \quad
\left\{\begin{array}{l}
e_k \mbox{ is k-$th$ eigenvalue, } e_1 > e_2 > \ldots > e_d \\
V_k \mbox{ is k-$th$ eigenvector} \\
\end{array}\right.
\]
Single out leading term:
\[
\rho = \rho_L+\rho_S, \quad
\rho_L = e_1*V_1*V_1^+, \quad
\rho_S = \sum_{k=2}^d e_k*V_k*V_k^+
\]
Here $\rho_L$ is coherent part of density matrix and
$\rho_S$ is the rest (incoherent part).
To be of physical meaning, this decomposition must be {\it stable}
with respect to variations of density matrix elements.
This is so if eigenvalues are {\it well separated}
in comparison with errors in $\rho$ matrix:
$ |e_1-e_2| \gg \sigma(\rho_{ij}) $.
This is the case for $\pi^+\pi^-\pi^-$ where $e_1\sim 1$, $e_2\sim 0.1$.

  Extraction of largest eigenvalue has the following advantages.
By construction matrix $\rho_L$ has rank one, so phases are well defined.
It quantitatively uses information about coherence factors,
which is often ignored.
Practical experience shows that resonance structures tend
to concentrate in $\rho_L$ and leakage (see below) is suppressed in $\rho_L$.
Nevertheless, $\rho_S$ can contain different non-leading
exchanges, albeit it often contains garbage.
We can also note that if the wave is small in $\rho_L$, its phase with
respect to largest waves can not be measured.

  As a restriction this method
requires a lot of data for good fit with small errors.
It is not applicable if eigenvalues are not separated. In this case
sometimes a group of clustered eigenvalues can be extracted.
It is also not applicable if all eigenvalues except one are not statistically
significant, as it is the case for unnatural sector for
$\pi^+\pi^-\pi^-$ system.

  An example of separation of largest eigenvalue is present in
figure~\ref{fig-lev}. On sub-figures a) and b) one can see a huge difference
between wave $2^-\P0^+\,\rho\pi$ in the full density matrix and in the
largest eigenvalue. It was already noted \cite{daum,pippm} that at low $t'$
region this wave is large and highly incoherent with others at
$M_{3\pi} \approx 1.2-1.4$~\mGeV{} and have only a relatively small shoulder
at $M_{3\pi}\approx 1.6$~\mGeV, which corresponds to well known $\pi_2(1670)$.
Our analysis confirms this as shown in figure~\ref{fig-lev}~a). Contrary to
this in the largest eigenvalue this wave is dominated by physical
$\pi_2(1670)$. The bump at low $M_{3\pi}$ is at least ten times suppressed,
which is consistent with its low coherence with other waves.
The physical nature of this phenomena is still unknown.

  In figures~\ref{fig-lev}~c),~d) one can see the difference between full
density matrix and largest eigenvalue for the wave  $4^+\G1^+\,\rho\pi$
at $|t'|<0.03$~\tGeV.
At both figures we can see at $M_{3\pi}\approx 2.0$~\mGeV{} a signal for 
$a_4(2050)$ while in fig.\ref{fig-lev}~c) we can also see a bump at
$M_{3\pi}\approx 1.3$~\mGeV{} which is absent in fig.\ref{fig-lev}~d).
This bump is a leakage from $a_2(1320)$ with intensity about 0.2\% of
total number of events or about 7.5\% of $2^+$ wave in this
low $|t'|$ region. This leakage is at least ten times suppressed in
coherent part of density matrix.

\section{Leakage study.}

\begin{figure}[tb]
\includegraphics[width=\textwidth]{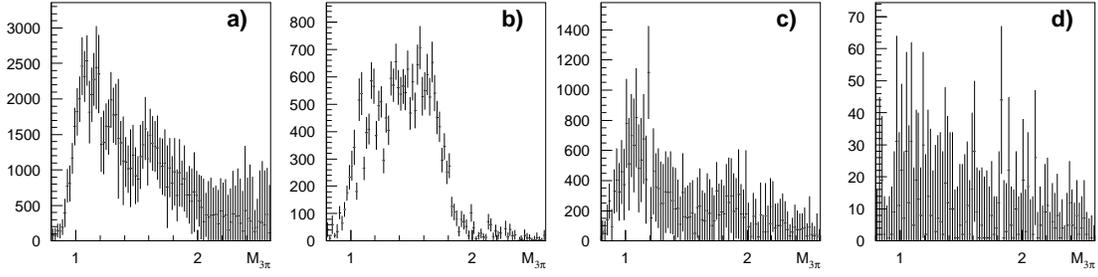}
\caption{Leakage study. Wave $1^{-+}\rho\pi$ at $0.03<|t'|<1.0$~\tGeV{}
in: a) real data, full $\rho$ matrix; b) real data, largest eigenvalue;
c) leakage, full $\rho$ matrix; d) leakage, largest eigenvalue.}
\label{fig-leak}
\end{figure}

  We have used the following method to study a possible leakage
effects due to finite setup resolution and limited knowledge of
setup acceptance. At first, we
fit real data with small but representative wave set ---
12 largest waves. The result of this step is a reasonably accurate
representation of multidimensional distribution of real events.
Next we generate Monte-Carlo events according to density matrix
from this fit, smear these events according to modeled setup resolution
and fit them as usual using standard wave set with all 42 waves.
To study dependence of results of variation of modeled acceptance,
we have used Monte-Carlo events without smearing,
but used corrupted MC program (with hodoscope trigger logic excluded) for
final fits with 42 waves.

  The results of leakage study for exotic wave $1^{-+}\rho\pi$ are shown
in fig.~\ref{fig-leak}. One can see that exotic wave can contain
30--50\% of leakage, but can not be described by leakage.
The structure of density matrix in the real data and leakage is quite
different --- in the coherent part of $\rho$ leakage is 20--50 times
suppressed. Most other waves can contain 5--10\% of leakage.

\section{Waves with $J^{PC}=1^{-+}$.}

\begin{figure}[t]
\includegraphics[width=\textwidth]{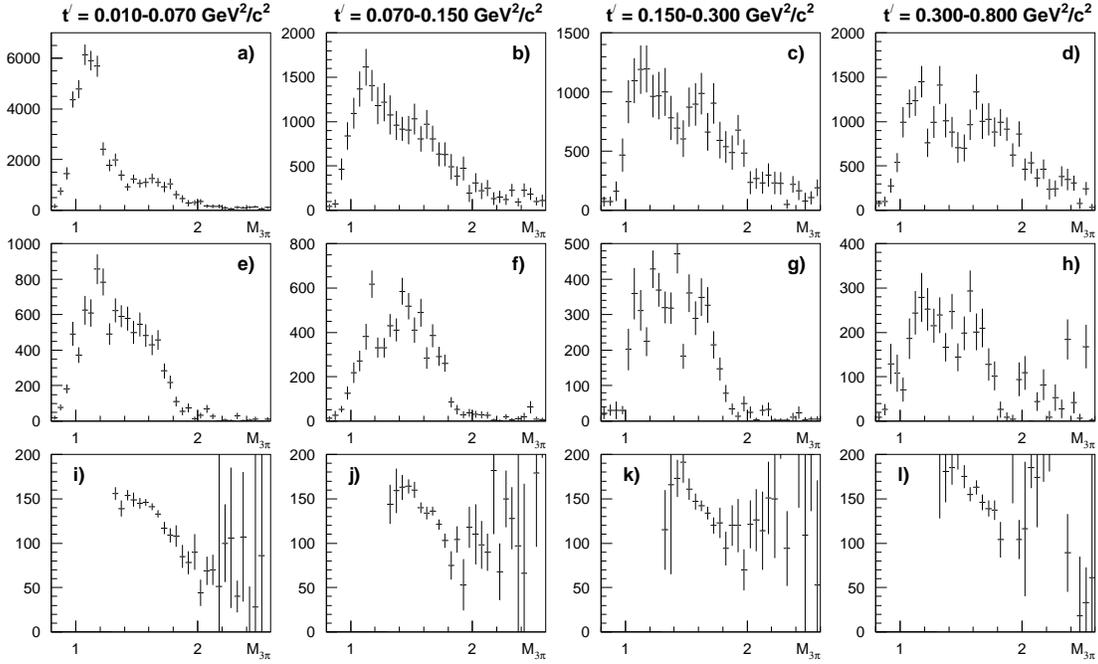}
\caption{Wave $1^-\P1^+\rho\pi$ at different $|t'|$ regions:
a--d) full density matrix; e--h) largest eigenvalue;
i--l) phase difference $\varphi(1^{-+}\rho\pi) - \varphi(2^{-+}f_2\pi)$,
full density matrix.}
\label{fig-onemp}
\end{figure}
  Waves with exotic quantum numbers $J^{PC}=1^{-+} \rho\pi$ were included
in our analysis with all possible projections $M^\eta = 1^+,\,0^-,\,1^-$.
Waves with $M^\eta = 0^-,\,1^-$ are small in comparison with
$M^\eta = 1^+$ (except for $M_{3\pi}<1.2$~\mGeV) and are not considered
by us as significant. Wave $1^-\P1^+\rho\pi$ in four different $t'$ regions
is shown in fig.~\ref{fig-onemp}. In the full density matrix
(fig.~\ref{fig-onemp}~a--d) at low $t'$ this wave consists mainly
of a bump around $M=1.0-1.2$~\mGeV{} and a shoulder at $M\approx1.6$~\mGeV{}.
At higher $t'$ regions the bump diminishes while the shoulder remains
approximately the same. From figures \ref{fig-onemp}~e--h) one can see
that this shoulder corresponds mainly to the coherent part of the
density matrix, and this part remains stable over investigated $t'$
region. The bins over $t'$ are selected so that numbers of events in
$a_2(1320)$ peak are approximately the same for all bins, namely about
20000~events/50~MeV, so $t'$ distribution for coherent part of exotic wave
is roughly the same as for $a_2(1320)$ while its crossection is only
2--3\% of it. One can see a sharp drop on the coherent part of exotic wave
at $M_{3\pi}\approx1.8$~\mGeV{}.
We can see analogous effect in some other
$\rho\pi$ waves and it can be connected with worse description of
high $M_{3\pi}$ region. Phase difference
$\varphi\bigl(1^-\P1^+\rho(770)\pi\bigr) -
 \varphi\bigl(2^-\S0^+f_2(1270)\pi\bigr)$
is shown in fig.~\ref{fig-onemp}~i--l). This phase difference is not
constant, visible drop corresponds to phase raise of $\pi_2(1670)$
resonance. Again the shape of phase variation is more or less stable
over inspected $t'$ region. In general we can not see here narrow
exotic object with $M\approx1.6$~\mGeV.

\section{Waves with $J^{PC}=0^{-+}$.}
\begin{figure}[t]
\includegraphics[width=\textwidth]{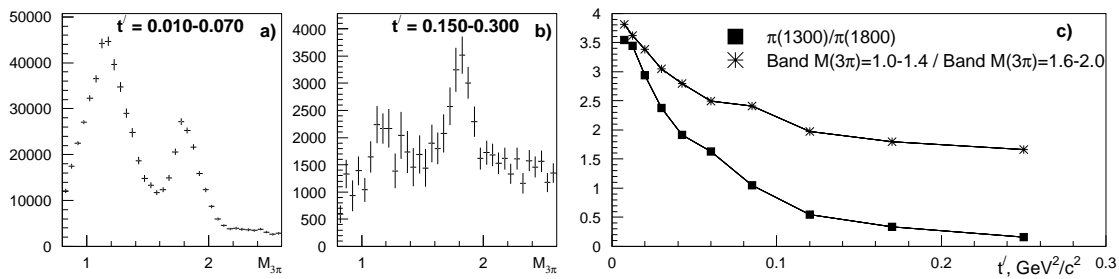}
\caption{Wave $J^{PC}=0^{-+}\eps\pi$: a) $|t'|=0.01-0.07$~\tGeV{} region;
b) $|t'|=0.15-0.30$~\tGeV{} region; c) relative $|t'|$ dependence of
$\pi(1300)$/$\pi(1800)$ signals and corresponding $M_{3\pi}$ bands.}
\label{fig-zeromp}
\end{figure}

\vspace{-2mm}
Results of PWA for the wave $0^-\S0^+\eps\pi$ in different $t'$ regions
are presented in figure~\ref{fig-zeromp}. At low $t'$ two peaks are
visible which corresponds to $\pi(1300)$ and $\pi(1800)$. At higher $t'$ peak
for $\pi(1300)$ is clearly suppressed. Relative strength of $\pi(1300)$ and
$\pi(1800)$ signals together with relative number of events in
$M_{3\pi}=1.0-1.4$~\mGeV{} and $M_{3\pi}=1.6-2.0$~\mGeV{} bands is shown
in fig~\ref{fig-zeromp}\,c) in ten $t'$ regions. One can see that
$t'$ distribution for $\pi(1300)$ is much more steep than for
$\pi(1800)$ or even for the total number of events in corresponding
$M_{3\pi}$ band. A possible explanation of this
phenomena can be given if the $\pi(1300)$ bump is at least partially 
non-resonant and produced by Deck-type final state scattering.

\vspace{-4mm}
\section{Conclusions.}
\vspace{-2mm}
  Partial wave analysis of $\pi^+\pi^-\pi^-$ final state at different
 $t'$ regions was performed on VES data. New type of amplitude analysis
was suggested --- extraction of the coherent part of density matrix.
Its advantages and limitations were briefly discussed.

  Wave $J^{PC}=1^{-+}\,\rho\pi$ was studied in different $t'$ regions
$t' =$~0.010--0.070--0.150--0.300--0.800 \tGeV. Wave shape is broad and more
or less the same in all $t'$ regions studied. Clear phase variation
with respect to $\pi_2(1670)$ is visible in all $t'$ regions.
No narrow object around $M=1.6$~\mGeV{} is found.

  Waves $J^{PC}=0^{-+}$ were studied in the same $t'$ regions.
Abnormally steep $t'$ distribution for $\pi(1300)$ was established.
This phenomena can be understood if $J^{PC}=0^{-+}$ wave in $\pi(1300)$
region is partially consists of Deck-type background.

\medskip
  This work is supported in part by INTAS-RFBR 97-02-71017,
RFBR 00-02-16555, RFBR 00-15-96689 grants.

\end{document}